\documentclass[aps,pre,twocolumn,groupedaddress,showpacs]{revtex4}

\RequirePackage{graphicx}
\RequirePackage{latexsym}

\begin{document}

\title{A Monte-Carlo and Self-Consistent Field calculations of encapsulated
spherical polymer brushes.}

\date{\today}

\author{Juan J. Cerd\`{a}, Tom\'{a}s Sintes and Ra\'ul Toral \footnote{Email
addresses: J. J. Cerda: jcerda@imedea.uib.es; T. Sintes: tomas@imedea.uib.es; R.
Toral: raul@imedea.uib.es}}
\affiliation{Departament de F\'{\i}sica and IMEDEA (CSIC-UIB).\\
 Universitat de les Illes Balears.
07122 Palma de Mallorca, Spain.}
\date{\today}


\begin{abstract}
We present the results of extensive numerical self-consistent field (SCF) and
3-dimensional off-lattice Monte Carlo (MC) studies of a spherical brush
confined into a spherical cavity. The monomer density profile and the cavity
pressure have been measured in systems where curvature of the cavity has an
important effect on the polymer brush conformation. A direct comparison between
the SCF and MC methods reveals the SCF calculation to be a valuable alternative
to MC simulations in the case of free and softly compressed brushes. In the
case of strongly compressed systems we have proposed an extension of the Flory
theory for polymer solutions, whose predictions are found to be in good
agreement with the MC simulations and has the advantage of being
computationally inexpensive. 
In the range of high compressions, we have found the monomer volume fraction
$v$ to follow a scale relationship with the cavity pressure $P$, $P
\sim v^\alpha$. SCF calculations give $\alpha=2.15 \pm 0.05$, close to {\em des
Cloiseaux law ($\alpha=9/4$)}, whereas MC simulations lead to $\alpha=2.73 \pm
0.04$. 
We conclude that the higher value of $\alpha$ obtained with MC comes from the
monomer density correlations not included in the SCF formalism.
\end{abstract}

\maketitle

\section{Introduction}
		
The structure and dynamics of polymer chains terminally anchored
or end-grafted to a surface is a very interesting problem from both,
theoretical and experimental points of view, due to the inherent complexity of
having chain molecules located in constrained geometrical environments. These
polymer systems are relevant in many areas of polymer science and
technology\cite{1,2}, such as chromatography, surfactants, lubrication,
adhesion, stabilization of colloidal systems, or even as potential drug
carriers\cite{1a,1b,1f,1l}. Properties related to these systems are also of
interest in the synthesis of nanoparticles\cite{nano}; spherical
brushes formed by colloidal poly-(methyl methacrylate) spheres with a grafted
layer of  poly(12-hydroxy stearic acid) (PMMA-PHSA) have served as a model  to
emulate hard-spheres minimizing the effect of van der Waals forces 
\cite{ref19}. More recently, the use of the light-scattering  properties of
coated spherical particles has been found to be a useful  non-destructive way
to probe and size systems ranging from blood cells to paper whiteners
\cite{ref18}.

Former theoretical studies by de Gennes and Alexander\cite{ref7,ref8}, Semenov
\cite{ref9}, Milner-Witten-Cates\cite{ref10,ref11} and
Zhulina-Priamitsyn-Borisov\cite{zhu} focused on polymer
chains grafted onto planar surfaces, and their predictions have been well 
supported by  Monte Carlo\cite{ref12,ref13,ref14,ref15} and molecular
dynamics\cite{ref16} simulations. 

The introduction of curved interfaces result in polymer brush structures 
whose properties differ significantly from those expected from flat interfaces.
Whereas the problem of concave surfaces, in which the surface curves towards
the polymer chain, was solved by Semenov\cite{8}, 
for convex surfaces it has not been possible to find a potential form that provides
a self-consistent solution leading to physical solutions. In this case, the
increased volume available to the stretched polymer, as it moves away from the
interface, is responsible of a rich physical behavior. For instance, colloidal
silica spheres with grafted alkane chains undergo a sol-gel transition when
dispersed in hexadecane solution\cite{grant}.

On the other hand, experimental studies on the rheological properties of
spherical  brushes appear rather scarce, certainly due to the strong
difficulties in the control of the grafting  procedure\cite{ref25}.
Nonetheless, several studies using SANS \cite{6c,6d}, ESR \cite{2k,6f} or NMR
\cite{6g,6h} technics have dealt with the physical chain-properties of
terminally attached homopolymers onto colloidal particles.

>From the theoretical point of view, there have been several attempts to
determine the interaction between spherical brushes. Daoud-Cotton\cite{ref20}
and Witten-Pincus\cite{ref21} in the limiting case of star shaped polymers;
Borukhov and Leibler\cite{ref22} using the Derjaguin approximation valid for
small curvature effects and short polymer chains.
Several numerical studies have been devoted to the study the properties of
single non-interacting brushes\cite{25,30,30a} and the interaction between two
spherical brushes\cite{Wijmans,CST}.

Whereas the previous studies have analysed the properties of a single or two
interacting spherical brushes, the understanding of the more complex and
interesting problem of a system of colloidal spherical brushes in solution 
remains untouched. 
Particularly, in the limiting case of high densities, the colloidal brushes are
subject to an isotropic pressure. In such case, the interaction of a colloidal
particle with the rest of the system can be modelled, in a first
approximation, by a single spherical brush confined inside a spherical
cavity. This model has the advantage of reducing the high computer cost.
On the other hand, this model turns out to be relevant to the study of the properties
of encapsulated dendrimers, liposomes, vesicles containing nanoparticles with
grafted chains, and might be of relevance in the synthesis of polymer-grafted
metal nanoclusters inside small material cavities or molecular cages.

The purpose of the present paper is to present the results of extensive MC and
SCF simulations of a single colloidal brush confined inside a spherical cavity
wall of variable radius. Polymer chains have an extent of the same order of the
core size of the colloidal particle, such that curvature effects are 
important. We have measured the monomer density profile and the cavity pressure
at different cavity radius. The results using MC and SCF methods have been
compared at low and high compression regimes, discussing the advantages of each
method.

The rest of the paper is organized as follows: in section II  we describe the
numerical procedures used to compute the monomer density profile and the cavity
pressure; in section III we present a detailed analysis of the numerical
calculations, this section also includes an extension of the Flory theory for a
free polymer solution, as an alternative way to obtain the pressure inside the
cavity; section IV concludes with a brief summary and a discussion of the
results.

\section{Numerical models}

\subsection{Monte Carlo method}           

\begin{figure}[htp]
\includegraphics[height = 5 cm, width = 5 cm]{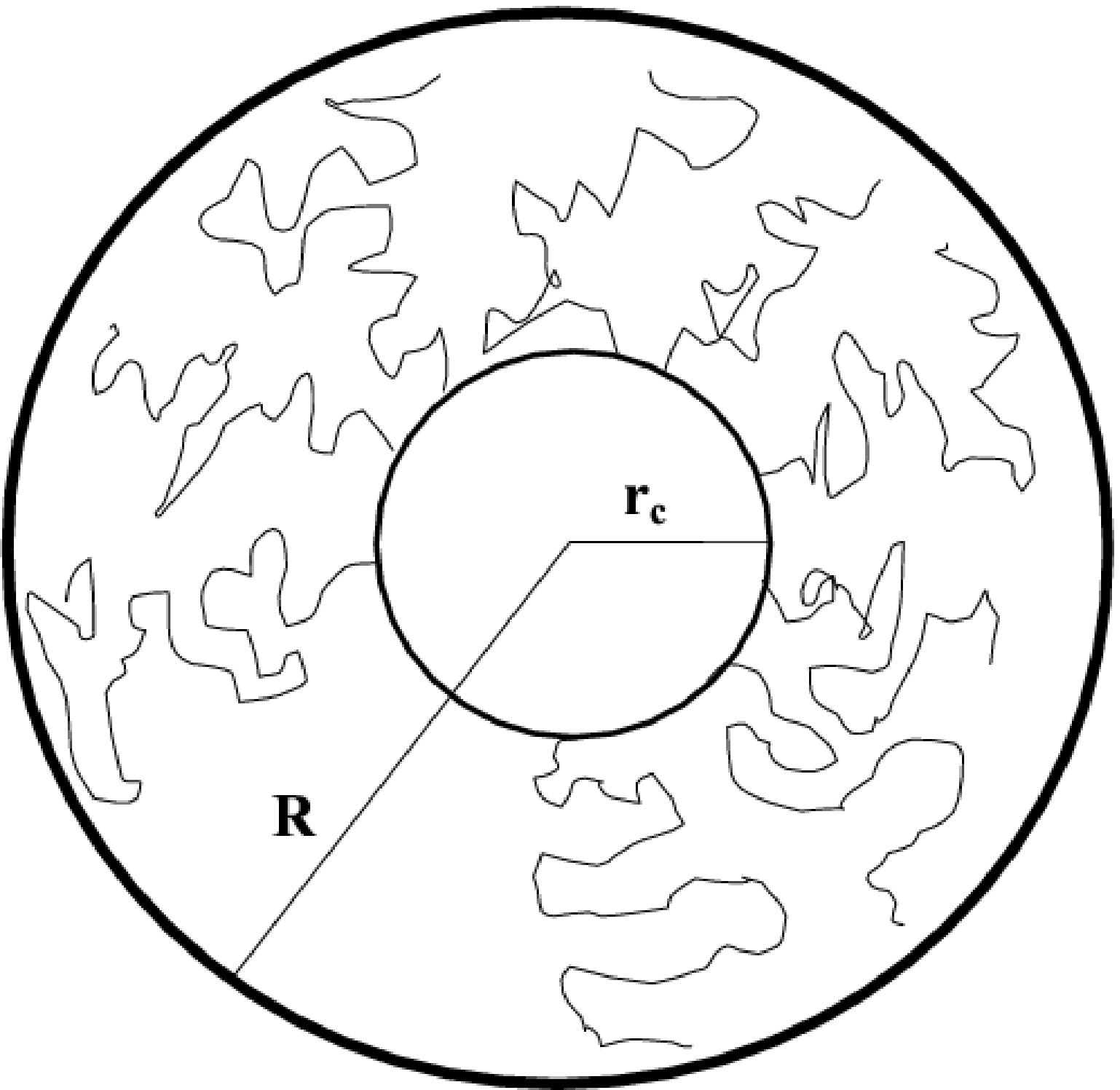}
\caption{Schematic representation of a spherical brush with an
impenetrable core of radius $r_c$ inside a spherical 
cavity of radius $R$.}
\label{figura1}
\end{figure}

In order to simulate the interaction between a spherical brush confined inside
a spherical cavity wall, we have used a 3-dimensional off-lattice Monte Carlo
method (MC).  We have generated the brush by homogeneously distributing  $f$
polymer chains grafted onto an impenetrable spherical surface of radius $r_c$.
The cavity wall is also impenetrable with a variable radius $R$. The polymer
chain is represented by the pearl necklace model\cite{29} containing $N$ beads
of diameter $\sigma$. The distance between two consecutive beads in the chain
is set to $1.1\sigma$. In all the simulations we have set $\sigma=1$. The
initial configuration of the self-avoiding polymer is randomly generated being
the first monomer permanently anchored to the surface (it is never allowed to
move). An schematic representation of the system is shown in Figure
\ref{figura1}. 

Monomers interact through a steric hard-core potential of the form:

\begin{equation} 
U_{steric}=\sum_{i,j=1}^{N \times f}V(r_{ij}),
\label{esint} 
\end{equation} 
where $V$ is a hard sphere potential:
\begin{equation} 
V(r_{ij})=\left\{
\begin{array}{l}
  0  ~~ \mbox{for  $|{\bf r}_i-{\bf r}_j| > \sigma$ ,} \\ 
 \infty ~ \mbox{for  $|{\bf r}_i-{\bf r}_j| <\sigma$.} 
\end{array} 
\right. 
\label{pote} 
\end{equation}

Different polymer configurations are generated by changing the position of a
randomly selected monomer. If the monomer is located inside the chain (between
the first monomer, permanently anchored, and the last one) its position changes
by rotating an arbitrary angle between $0$ and $2\pi$ around the axis 
connecting the previous and following monomers in the chain. Chain ends just
perform random wiggling motions. The proposed motion is accepted if the
excluded volume interaction is preserved (Eq. \ref{pote}). A link-cell method
\cite{link}  has been implemented in the algorithm to efficiently check all
possible monomer overlaps. 

Initially, the radius of the cavity wall is set to be larger than the usual
extent of the brush to ensure that during the equilibration process no
interaction between the brush and the cavity wall occurs.  We define one Monte
Carlo Step (MCS) as $ N \times f$ trials to perform monomer moves.  The
spherical brush has been equilibrated typically during  $5\times 10^5$ MCS.
After this initial equilibration time, magnitudes of interest are recorded
every $10$ MCS. In what follows, the expression for the force and interacting
potential are given in units of $k_BT$.

The force that the system exerts to avoid compression is computed as the change
in the  free energy  ${\cal F}$ due to an infinitesimal change in the cavity
radius $R$ is given by:
\begin{equation}
F(R) \equiv \frac{\partial \ln Z(R)}{\partial R},
\end{equation}
where $Z(R)=\exp(-{\cal F}/k_BT)$ is the partition function of the system. 
Due to the hard--core structure of the potential, see Eq.(\ref{pote}), the 
partition function 
\begin{equation}
Z(R) =  \int{ \prod_{i=1}^{N \times f} d{\bf r_i}~~ 
        \exp\left(-\sum_{j=1}^{N \times f}  
        V({\bf r_{ij}}) \right) }
\label{eqzr}
\end{equation}
is equal to the volume $\Omega(R)$ of all possible polymer chains 
configurations compatible with the steric requirements.

Let $\Omega_C(R) \subset \Omega(R)$ be the subset of configurations
compatible  with a reduction of the radius cavity by an amount  $\delta R$.
Similarly, let  $\Omega_E(R-\delta R) \subseteq \Omega(R-\delta R)$  be the set
of polymer configurations of $\Omega(R-\delta R)$ in which it is  possible to
increase the radius of the cavity by an amount $\delta R$.  The compression
probability is just  $P_C(R)=\Omega_C(R)/\Omega(R)$, and the probability that
an expansion can be done is $P_E(r-\delta R)=\Omega_E(R-\delta
R)/\Omega(R-\delta R)$. Since no steric requirement can 
prevent the expansion of the  cavity, $P_E(R)=1$ for all values of $R$.
The subsets $\Omega_C(R)$ and $\Omega_E(R-\delta R)$ are in one to one
correspondence, consequently  
\begin{equation} 
\frac{Z(R)}{Z(R-\delta R)} = \frac{\Omega(R)}{\Omega(R-\delta R)} = 
                             \frac{P_E(R-\delta R)}{P_C(R)} = \frac{1}{P_C(R)}
\label{f0} 
\end{equation}

Because of the finite step $\delta R$ used in computer simulations, we
approximate 
\begin{equation}
\frac{\partial \ln Z(R)}{\partial R} \approx
\frac{1}{\delta R} \ln \frac{Z(R)}{Z(R-\delta R)}=
\frac{-1}{\delta R} \ln {P_C(R)}
\label{f00}
\end{equation}

The compression probabilities are obtained by spanning directly this 
probability for a given cavity size $R$. Then, we force to compress the system
to a new cavity size $R- \delta R$. The system is equilibrated again during 
$10^5$ MCS before new measurements are taken.

The monomer radial density $\phi$ is defined as usual:
\begin{equation}
\phi(r)=\frac{\nu(r)}{4 \pi r^2 dr},
\label{conc}
\end{equation}
being $\nu(r)$ the number of monomers located within a distance between $r$ 
and $r+dr$ from the centre of the sphere. 
The definition of  ${\phi}(r)$ is such that the following normalization holds:
\begin{eqnarray}
~{\int}_{0}^{\infty}d^3r ~ {\phi}(r) = Nf 
\end{eqnarray}
In a similar way, we also define the chain-ends monomer density,
${\epsilon}(r)$, where $\nu(r)$ takes only into account the number of free
chain ends.

\subsection{Self Consistent Field method}           
                  
In order to compute the  probability density function (pdf) for polymer
systems, it is customary to deal with a Schr\"odinger-like equation  for the
pdf for a single chain\cite{7e}. The use of this formalism is a valid
approximation  for spatial scales much larger than the polymer blob size. 
However, for encapsulated polymer brushes, and mainly at moderate and high
compressions, the spatial scale of the cavity is comparable to the blob size.
Thus, the use of the precedent formalism to compute the pdf becomes
inadequate. Instead, we compute the pdf of a polymer chain  using directly
the recurrence law for the pdf from which the Schr\"odinger-like
equation derives\cite{7e}. We must be aware that the use of the recurrence law still
implies some approximations. It is assumed that on each monomer of the polymer
chain acts a potential that only depends on the position of the monomer in the
system, $U=U({\bf r})$, therefore, bond correlations are not taken into
account.  It is also assumed that the potential is a function of the local 
monomer density, $U(\phi ({\bf r}))$, ignoring particle density correlations. 
Finally, the properties of the whole ensemble of chains are deduced from the
pdf of a single chain.

Under the above assumptions, the spherical
cavity is discretized  in concentrical shells of thickness $dr$ and all
monomers inside a shell are assumed to be equivalent. 
A polymer chain composed by $N$ monomers is represented as a path of $N$
segments of length $\sigma$. Each segment is labeled by an index $\tau$
associated to the spherical shell at which it belongs.
The pdf associated to all possible paths composed by $n$ segments, being the
first segment inside the shell ${\bf h^{\prime} }$ and the last one inside the
shell ${\bf h}$, is defined as
\begin{eqnarray}
G_{n}({\bf h^{\prime}},{\bf h}) &\equiv & 
\sum{}^{A}~{\mathrm e}^{-\sum{}_{i=1}^{n} ~ U({\bf  {\tau_i}})} 
\end{eqnarray}
in which $\sum{}^{A}$  stands for a sum over all the hypothetical $n$-paths
that join the shells ${\bf h^{\prime}}$ and ${\bf h}$.  
This function verifies
$G_{n}({\bf h^{\prime}},{\bf h}) =  G_{n}({\bf h},{\bf h^{\prime}}) $. 
Therefore, the pdf associated to a path of $n+1$ segments may be written as 
\begin{eqnarray}
G_{n+1}({\bf h^{\prime}},{\bf h}) &=&  \sum{}^{A}~\left(
{\mathrm e}^{-\sum{}_{i=1}^{n} ~U({\bf  {\tau_i}})} ~~
{\mathrm e}^{-U({\bf  {\tau_{n+1}}})} \right)
\end{eqnarray}
Assuming $U({\bf  {\tau_{n+1}}})$ to be independent of rest of the
segments in the chain and of the starting point of the sequence, the
precedent equation reads: 
\begin{eqnarray}
G_{n+1}({\bf h^{\prime}},{\bf h}) &=& \left( \sum_{{\bf h^{\prime \prime}}}
{}^{D}~ G_{n}({\bf h^{\prime}},{\bf h^{\prime \prime}})
\right) ~~{\mathrm e}^{-U({\bf h})}
\end{eqnarray}
where $\sum{}^{D}$ implies a sum over all the ${\bf h^{\prime \prime}}$  shells
from which we can get into shell ${\bf h}$ using a single segment, therefore,
shells ${\bf h^{\prime \prime}}$ and ${\bf h}$ are at a relative distance less
or equal to $\sigma$. The above equation stands for the recurrence law needed
to calculate the chain pdf once the potential $U({\bf h})$ is given.

We have set the interaction potential $U(i)$ to be proportional to the monomer 
concentration in shell $i$, ${\phi}(i)$:
\begin{eqnarray}
U(i) &=& {\omega} ~ {\phi}(i)
\label{uphi}
\end{eqnarray}
where $w$ is the excluded volume parameter defined as\cite{4a}:
\begin{eqnarray}
{\omega} = 4{\pi}{\int}_{0}^{\infty} (1 - {\mathrm e}^{-V(r)}) r^{2}dr
\end{eqnarray}
where $V(r)$ is the interaction potential introduced in Equation (\ref{pote}),
that for a value of $\sigma=1$, as in the MC method, we obtain 
$\omega= 4 \pi/3 \approx 4.2$

The monomer concentration ${\phi}(i)$ and the free-end-chain 
concentration  ${\epsilon}(i)$  are defined as:
\begin{eqnarray}
{\phi}(i)=\frac{f}{4{\pi}r_i^2dr} 
\frac{{\mathrm e}^{U(i)}~ 
{\sum}_{n=0}^{n=N}~{\sum}_{j=i_{r_c}}^{i_R} G_n(i_{r_c},i) G_{N-n}(i,j)  }
{{\sum}_{j=i_{r_c}}^{i_R}G_N(i_{r_c},j) }
\end{eqnarray}

\begin{eqnarray}
{\epsilon}(i)=\frac{f}{4{\pi}r_i^2dr} 
\frac{ G_{N}(i_{r_c},i)  }
{{\sum}_{j=i_{r_c}}^{i_R}G_N(i_{r_c},j) }
\end{eqnarray}
where $f$ is the number of chains and $i_{r_c}$ and $i_R$ are the shell indexes
with radius equal to the brush core surface and the cavity wall respectively.
The factor ${\mathrm e}^{U(i)}$ in the monomer concentration is introduced 
to avoid double counting of the interaction term ${\mathrm e}^{-U(i)}$ at
shell $i$ that comes from splitting the pdf into two terms, one that ends at
shell $i$ and the other starting at the same shell. 

The change in the  free energy when we reduce the cavity size, 
${\Delta}{\cal F}(i_R)$, is given by
\begin{eqnarray}
{\Delta}{\cal F}(i_R) &=& {\ln} \left( \frac{ {\Omega}(i_R)}
{{\Omega}({\infty})} \right)
\label{scff1}
\end{eqnarray}
being
\begin{eqnarray}
{\Omega}(i) &=&f ~ {\sum}_{j=i_{r_c}}^{i}G_N(i_{r_c},j).
\end{eqnarray}
We have set the reference state as a system with a cavity wall at a distance
far enough so that no chain can reach this wall. Therefore, the force to
compress the brush, given a cavity size $R$, is: 
\begin{eqnarray}
F(R) &=& -  \frac{ \partial{ ({\Delta}{\cal F}(R) )}}  {\partial{R}}
\label{scff2} 
\end{eqnarray}

Given R, N and f, an iterative process is used to obtain the pdf and
the density profiles. We have iterated the process until self-consistency is
reached. We set the condition for self-consistency such that the sum of the
square differences in the density profiles coming from two consecutive
iterative steps is less than $10^{-8}$.

The SCF method has the great advantage  of being three to four orders of
magnitude less expensive in computer time than the MC method.
In the next section we will show that SCF and MC calculations give similar
results for the density profiles $\phi(r)$ and $\epsilon(r)$ in the case of
free polymer brushes. However, SCF predictions for the cavity pressure worsen
for highly compressed systems.

\section{Results and discussion}

\subsection{Density profiles}

We have performed extensive numerical calculations for free and encapsulated
spherical polymer brushes in order to computed the monomer density profiles and
the compression forces for different sets of parameters $(R,N,f)$.
The core radius of the colloidal particle where polymers are grafted is taken
to be $r_c=5\sigma$ and is kept constant through all the simulations. The
diameter of the monomers is set to $\sigma=1$.  We have taken polymer chain
lengths in the range of $N=30$ to $N=70$, and we have varied the number of
grafted chains from $f=5$ to $f=75$. The range of parameters $(N,f)$ has been
chosen in order to obtain chain extents roughly of the same order than the 
diameter of the core where curvature effects are important. For the SCF method
we have used a shell thickness $dr=0.1\sigma$.

\begin{figure}[H]
\includegraphics[height = 6 cm, width = 8 cm,angle=0]{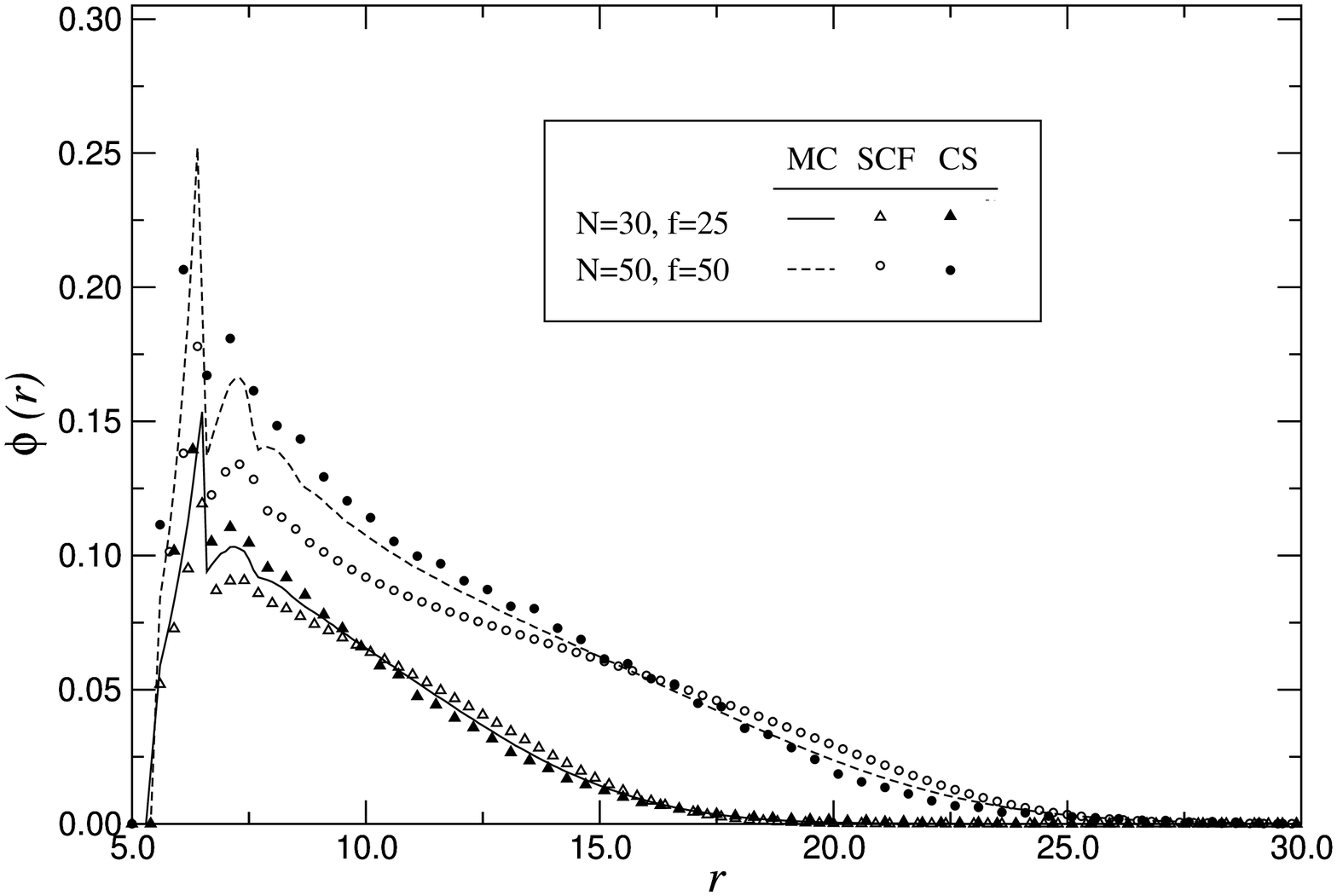}
\caption{Comparison between the monomer density profile $\phi(r)$ for free or
uncompressed spherical polymer brushes obtained from our MC simulations (lines)
and SCF calculations (symbols) for different values of the chain length $N$ and
number of grafted chains $f$. The results are compared to the predictions of
Cariagno and Szleifer\cite{Carig} (CS, filled symbols)}
\label{figura2}
\end{figure}

We have first studied the monomer density profiles $\phi (r)$ for an
unconstrained spherical brush using our MC simulations and the SCF formalism. 
In Figure \ref{figura2} we present the results for two different values of the
chain length $N$ and number of grafted chains $f$ that correspond to
representative values for soft and densely packed brushes. We can observe a
good agreement between the MC and SCF calculations. The density oscillations
observed at small $r$, close to the core of the colloidal particle, are
originated due to wall-effects of the impenetrable core. For a free spherical
polymer brush Cariagno and Szleifer\cite{Carig} computed the monomer density
profile derived from a single-chain mean field theory. A comparison between the
Cariagno and Szleifer data (CS) and our results is also included in Figure
\ref{figura2}. The better agreement of the CS predictions with our MC
simulations, in contrast to the SCF calculations, can be understood in the
sense that the Cariagno and Szleifer formalism requires a representative sample
of chain configurations as an input data to solve the equations that we have
generated using our MC method.

\begin{figure}[H]
\includegraphics[height = 18 cm, width = 8 cm,angle=0]{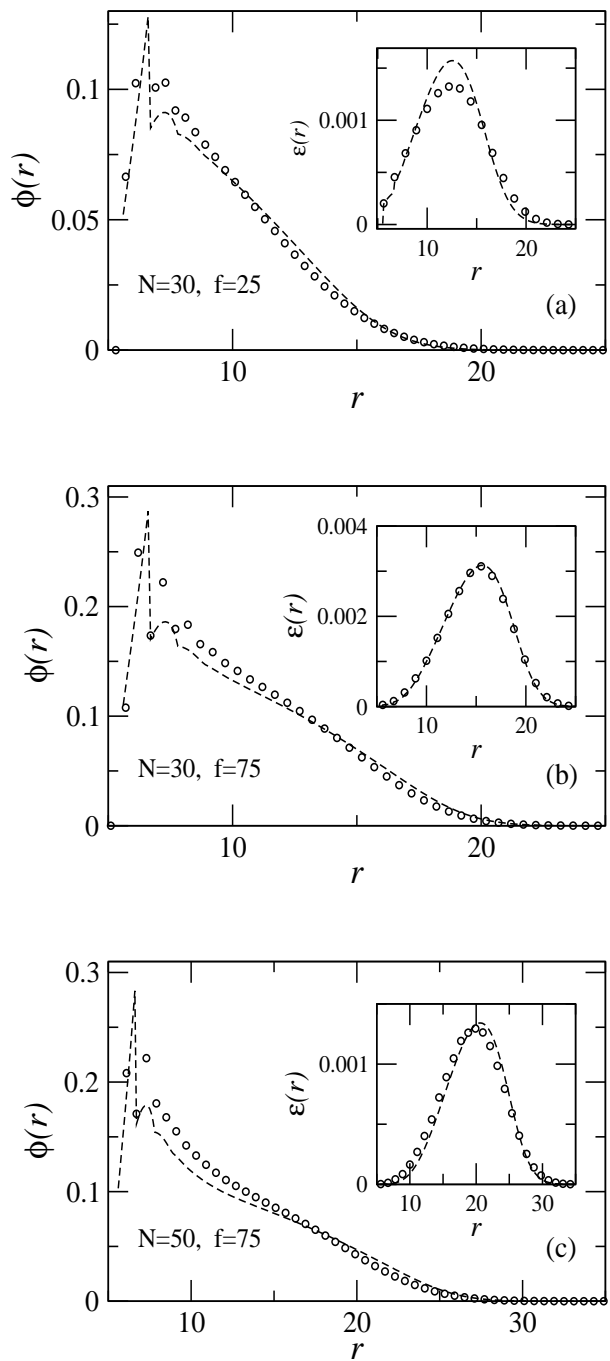}
\caption{Comparison between the monomer density profiles ${\phi}(r)$ for
uncompressed spherical brushes obtained with MC ($\circ$) and SCF (dashed
lines) calculations. Inset: chain--end density profiles. From top to bottom:
(a) $N=30, f=25$; (b) $N=30, f=75$; (c) $N=50, f=75$.}
\label{figura3}
\end{figure}

In Figure \ref{figura3} we compare the monomer and chain--end density
profiles for unconstrained polymer brushes obtained from our MC and SCF
calculations. Figure \ref{figura3}(a) stands for $N$=30 and $f$=25, whereas
Figure \ref{figura3}(b) and Figure \ref{figura3}(c) show the results for
($N$=30,$f$=75) and ($N$=50,$f$=75) respectively. In all cases the profiles are
roughly similar. In particular, both methods agree very well in the case of the
chain-end density profile $\epsilon(r)$ for short chains in a densely packed
brush (see inset of Figure\ref{figura3}(b)). This is mainly due to the fact
that the chains are forced to be mostly fully stretched out and density
correlations, not present in the SCF formalism, are not relevant. On the other
hand, the results for the monomer density profiles $\phi(r)$ show systematic
differences, although small, between the MC and SCF calculations. Close to the
core of the colloidal brush, SCF results display density profiles slightly
smaller than those obtained via MC simulations. And vice versa, in an
intermediate region, the SCF method gives densities slightly larger than in the
MC simulations. 
The same systematic behaviour was found by  Cosgrove et al.\cite{3e}, when
comparing MC and SCF density  profiles for flat brushes. Cosgrove attributed
these differences to the fact that MC simulations accounts explicitly for the
excluded volume effect, whereas SCF accounts only approximately for this effect.

\subsection{Cavity pressure and force profiles}

We have measured the force profile exerted by an encapsulated spherical polymer
brush onto the external cavity wall through the evaluation of the changes in
the free energy due to an infinitesimal change in the radius of the cavity. Within
the MC simulations, the force can be calculated by directly measuring the
compression probabilities (Equation \ref{f00}); whereas in the SCF approach,
once we have reached self-consistency, we use the pdf in Equations \ref{scff1}
to \ref{scff2}. 

Alternatively, we propose another way to compute the force profile and the
cavity pressure that is derived from the Flory theory for polymer
solutions\cite{4h}. This method has the advantage of being computationally
inexpensive, and their predictions will be compared with the MC and SCF
calculations.

\subsubsection*{An extension of the Flory theory for polymer solutions}

In the Flory theory the osmotic pressure in a polymer solution can be written
as:
\begin{eqnarray}
{\Pi (R)} &=& \frac{-1}{{\cal V}} \left( 
{\ln}(1-v) + (1 - \frac{1}{N}) v + {\chi} v^{2} \right)
\end{eqnarray}
where ${\cal V}$ is the molar volume of the solvent, $v$ is the volume
fraction of  solute, $N$ is the degree of polymerization or chain length,
and ${\chi}$ is the  Flory parameter. We set ${\chi}=0$ which is the
condition of a dry-brush. Under this assumption, the contributions to the free
energy come only from the entropy associated to all possible configurations of
the system. We suppose that the volume of the solvent and the volume fraction
of the solute are given respectively by:
\begin{eqnarray}
{{\cal V}} &{\sim}& ({\tau} - aNf )\\
v &=& \frac{aNf}{\tau}
\label{vfr}          
\end{eqnarray}
$a$ is the volume of a single monomer. $\tau$ is the total available cavity
volume between the inner wall, represented by the core of the colloidal
particle where chains are grafted and the  cavity wall, thus:
\begin{eqnarray}
{\tau (R)} &=& \frac{4{\pi}}{3}(R^{3}-r^{3}_{c})
\end{eqnarray}

At variance with the original Flory treatment, the molar volume of the
solvent ${\cal V}$ refers to the remaining space in the system once we
have subtracted the volume occupied by the monomers, thus, it is no longer a
constant value.

The force to compress  the cavity will be proportional to the area of the
cavity wall and to the change in the osmotic  pressure, thus:
\begin{eqnarray}
F(R) & {\sim}  4{\pi}R^2 {\Delta}{\Pi}(R)
\label{flo1}
\end{eqnarray}

\begin{figure}[H]
\includegraphics[height = 18 cm, width = 8 cm,angle=0]{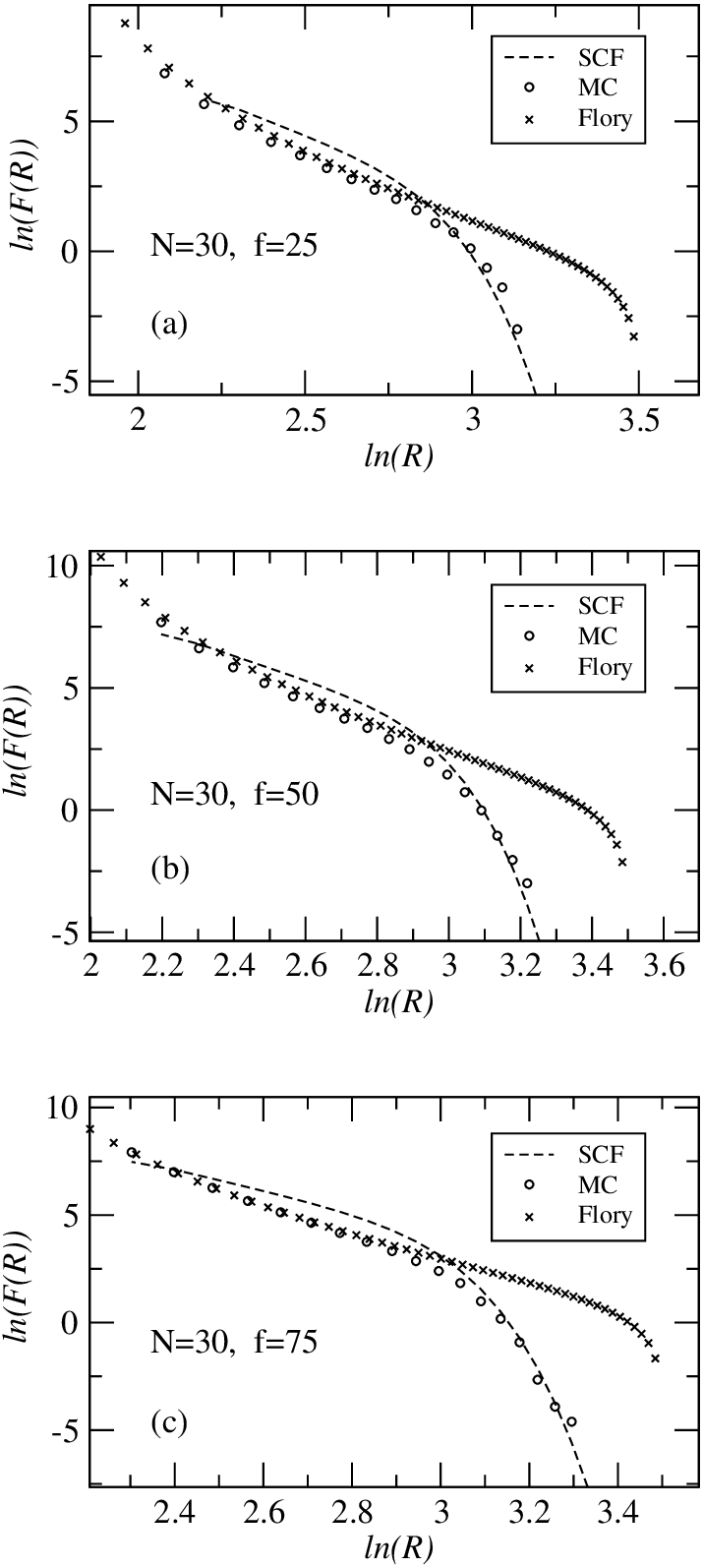}
\caption{Log-log plot of the force profile of an encapsulated spherical polymer
brush vs. the cavity size  $R$, for polymer chains of length $N=30$. MC results
are represented by filled circles; SCF data by dashed lines; and the
predictions coming from the Flory theory by crosses. Different figures stand
for different number of grafted chains $f$. From top to bottom: (a) $f=25$; (b)
$f=50$; (c) $f=75$.}
\label{figura4}
\end{figure}

\begin{figure}[htp]
\includegraphics[height = 6 cm, width = 8 cm,angle=0]{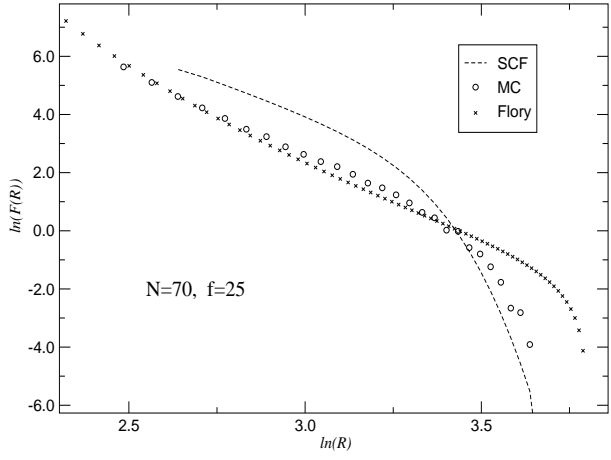}
\caption{Same as Figure \ref{figura4} for the single case of long polymer
brushes with $N=70$ and $f=25$.} 
\label{figura6}
\end{figure}

In Figures \ref{figura4} to \ref{figura6} we present in log-log plots the force
profile $F(R)$ vs. the cavity size $R$ computed for different values of the
polymer chain length $N$ and number of grafted chains $f$. In each figure we
include the results coming from the MC simulations (circles), the
predictions of the SCF theory (dashed lines) and results derived from the
application of the extended Flory theory (crosses). Figure \ref{figura4}
concentrates on the results derived for short polymer chains ($N=30$); Figure
\ref{figura5} for intermediate chain lengths ($N=50$); and Figure \ref{figura6}
for long polymers ($N=70$). In all the cases studied we have used the same
fitting constant to adjust the predictions of the extended Flory theory (see Eq.
\ref{flo1}), and we have taken, as a reference state, a cavity size $R^*$ at
which $\ln(F(R^*)) \rightarrow -{\infty}$ in the MC simulations. 

A direct comparison between SCF and MC force profiles shows a rather good
agreement for weakly compressed systems. However, systematic differences are
observed for intermediate and high compression values. In the intermediate
region we found the SCF forces to be larger than the ones derived from the MC
simulations, whereas for high compressions it is the MC force the one that
becomes larger than the SCF outcome. 

The mismatches observed between the SCF and MC results are due to a twofold
effect. For small cavity sizes or highly compressed systems, the assumption in
the SCF model of a linear dependence of the mean field potential with the
monomer density (see Eq. \ref{uphi}) breaks down. In fact, under this
assumption, the SCF formalism allows a cavity size smaller than the volume
occupied by the polymers without requiring an infinite force. On the other
hand, for intermediate compression values, the larger forces obtained with the
SCF formalism are originated in an overestimation of the monomer interactions. 
The SCF method does not include the effect of monomer correlations, thus it
allows higher average densities in the system than the ones found in the MC
simulations. As a consequence, stronger repulsions between the polymers take
place and a higher force is required to compress the brush.

The predictions of the extended Flory theory are found, remarkably, despite its
simplicity, to be in a very good agreement with the results of the MC
simulations in the intermediate and high compression regimes. However, the
force is overestimated for weakly compressed systems. This result is easily
explained since the Flory theory was formerly developed for free polymer
chains. As the cavity size grows, the difference between a system of grafted
chains and a polymer solution becomes evident and, as it is expected, the
interaction of the grafted chains with the outer surface is much weaker than
the one coming from a polymer solution.

It is worth to notice that the MC data can be fitted remarkably well with the
extended Flory theory for intermediate and high compression values, and with the
SCF formalism for weakly compressed systems.

\begin{figure}[htp]
\includegraphics[height = 6 cm, width = 8 cm,angle=0]{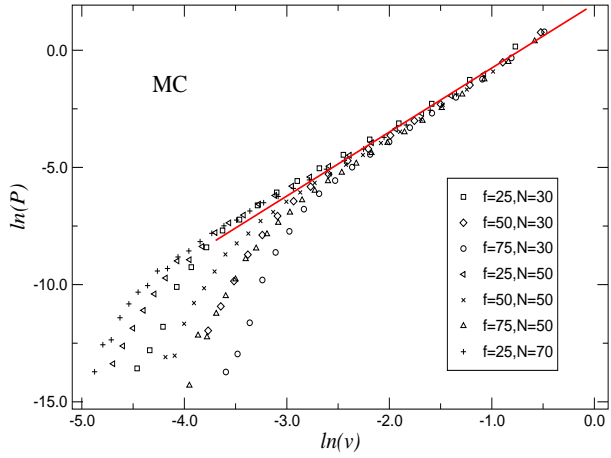}
\caption{Log-log plot of the system pressure $P$ vs. the monomer volume
fraction $v$ obtained from MC simulations for different values of the chain
length $N$ and number of grafted chains $f$.  A solid line of slope 2.73 is
included to guide the eye.}
\label{figura7}
\end{figure}

\begin{figure}[H]
\includegraphics[height = 6 cm, width = 8 cm,angle=0]{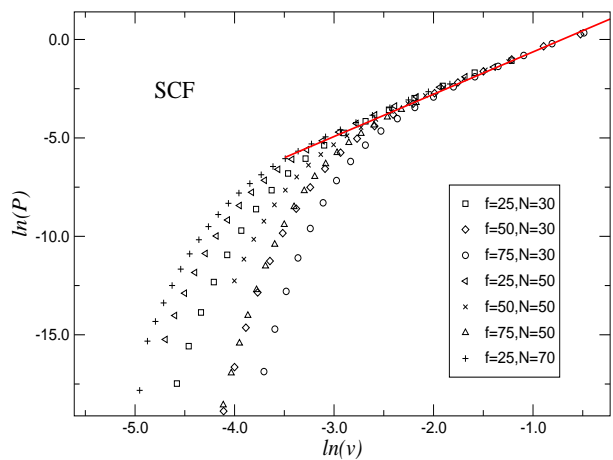}
\caption{Log-log plot of the system pressure vs. the monomer volume fraction
$v$ derived from the SCF formalism. A solid line of slope 2.15 is included to
guide the eye.}
\label{figura8}
\end{figure}

We have analysed the relationship between the monomer volume fraction $v$ (see
Eq. \ref{vfr}) and the pressure exerted on the cavity wall $P$ defined as the
force to compress the system divided by the area of the cavity.  The
results for the MC and SCF calculations are shown in log-log plots in Figure
\ref{figura7} and Figure \ref{figura8} respectively. In both cases, we find a
complex behavior of the monomer volume fraction with the cavity pressure for
weakly compressed systems that depends on the different values of the polymer
chain length $N$ and number of grafted chains $f$. But, for increasing values
of the cavity pressure $v$ and $P$ follow a power-law of the form $P \sim
v^{\alpha}$ independent of $N$ and $f$. The best fit to the numerical data gives
a slope of $\alpha= 2.73 \pm 0.04$ for the MC simulations and $\alpha= 2.15 \pm
0.05$ for the SCF results. The exponent obtained for the SCF data is very close
to the des Cloiseaux power law ($\alpha=9/4$) found in semi-dilute polymer
solutions \cite{7e}. For large monomer concentrations the polymer theory
predicts that all thermodynamic properties must reach values that are
independent of the degree of  polymerization, as we have observed in both MC
and SCF methods. We must be aware that des Cloiseaux law is deduced from
scaling arguments and neglects non-linear dependences on the concentration.
This fact might explain the agreement between SCF and des Cloiseaux law,
whereas MC results follows a power law which exponent is larger due to monomer
density correlation effects. 

Differences between force profiles derived from SCF formalisms, and those
obtained from other methods that account for chain interdigitation and
correlations between the nearest-neighboring bonds are also referenced in
several works. For instance, Ruckenstein-Li\cite{7c} (to be referred as RL)
compared the experimental force profile of two interacting crossed cylinders
bearing grafted polymer chains, with the numerical data obtained  with a
generator matrix formalism and from SCF methods. RL found the matrix formalism
to provide a better agreement with the experimental data than the SCF results. 
In most cases, the force profiles derived by RL with the matrix formalism and
the SCF method are rather similar to the ones we found comparing the MC
simulations and the SCF calculation. This reinforces our presumption that
interdigitation and monomer correlations are responsible of the observed
differences between the MC and SCF data. Additionally, the fact that mean field
calculations overcounts the segment-segment interactions in the evaluation of
the free energy, as mentioned above, has already been noticed by
Lin and Gast \cite{4f}.

\section{Summary and concluding remarks}

In this paper we have studied the behavior of an encapsulated spherical brush
inside a spherical cavity. This system is a first approximation to model the
interactions in a solution of colloidal particles bearing grafted polymer
chains onto their surface. This model is particularly relevant for moderate and
high density values where the colloidal brushes are subject to an isotropic
pressure, and might be of relevance to the behavior of encapsulated dendrimers,
liposomes or vesicles containing polymer brushes as well. 

We have measured the monomer density profile and the cavity forces through 
extensive 3-dimensional off-lattice Monte-Carlo simulations and using a
Self-Consistent Field formalism. In the latter case, we have used directly the
probability density function recurrence law for the  propagator
$G_N(r,r^{\prime})$, avoiding the length scale approximation involved in
self-consistent field methods that uses Schr\"odinger-like equations. 
Alternatively, we have proposed a theoretical description based on an extension
of the Flory theory for polymer solutions to compute the pressure inside the
cavity.

A comparison of the predicted forces exerted by the polymer brush onto the
cavity surface among the different methods reveals the following: 
$i$) For weakly compressed systems, MC and SCF data show a rather good
agreement. However, the force is overestimated in the extended Flory theory.
This difference arises since the Flory theory was developed for free
polymer chains and not for polymer brushes, thus, its prediction is not
physically relevant when the cavity wall is located at a distance larger than
the typical brush extension.
$ii$) For intermediate and highly compressed systems the MC data agrees
reasonably well with the results derived from the Flory theory. In the
intermediate regime, it is the SCF formalism that overestimate the force. This
behavior can be easily explained since the SCF method does not account for
monomer correlations, allowing higher monomer densities and thus, higher forces
are required to compress the brush. On the other hand, for highly compressed
systems, the linear dependence of the mean field potential with the monomer
density turns out to be inadequate (the repulsion between monomers allows a
volume reduction of the system beyond the own excluded volume of the monomers
at a finite energy cost) leading to lower forces than the ones derived from MC
simulations.

We have found a power law relationship between the monomer volume fraction and
the cavity pressure, $P \sim v^{\alpha}$. SCF data gives a slope of
$\alpha=2.15$, very close to the des Cloiseaux law derived for semi-dilute
polymer solutions. On the other hand, the MC simulations provide a larger
exponent $\alpha=2.73$ that is originated in the monomer correlations not
present in the previous models.

The study of a spherical brush constrained by an isotropic steric wall is
directly related to the problem of a dense solution of colloidal particles with
grafted chains onto their surface where chain  interdigitation among brushes is
not favoured. We expect these results  to stimulate further theoretical and
experimental studies towards the understanding of  the behaviour of colloidal
particle systems in solution. In this sense,  the computed force can be directly
related to the measure of disjoint pressures by using the expressions derived
by Evans and Napper\cite{EvNa}.

\section*{Acknowledgments}
Financial support from the Spanish MCyT grants nos. BFM2000-1108 and 
BFM2001-0341-C02-01 are acknowledged.

\newpage

\begin{figure}[H]
\includegraphics[height = 18 cm, width = 9 cm,angle=0]{figure5.eps}
\caption{Same as Figure \ref{figura4} for polymer chains of length $N=50$. From
top to bottom: (a) $f=25$; (b) $f=50$; (c) $f=75$.}
\label{figura5}
\end{figure}

\begin{thebibliography}{99}

\bibitem{1} Napper, D. H. {\em Polymeric Stabilization of colloidal
dispersion}; Academic, London, 1983.

\bibitem{2} Russel, W. B;  Saville, D.A. and  Schowalter, W.R. {\em Colloidal
Dispersions}; Cambridge University Press, Cambridge, 1989.  

\bibitem{1a}  Hsu, W. P.;  Yu R. and  Matijevic, E., J. of Colloid Interface Sci. 
{\bf 1993}, 156, 36.

\bibitem{1b} Meyer, R. A., Appl. Opt.,{\bf 1979}, 18, 585. 

\bibitem{1f} van Zanten, J. H.  and Monbouquette, H. G.,  J. Colloid Interface
Sci. {\bf 1991}, 146, 330.

\bibitem{1l} Kuhl, T. L. ; Leckband, D. E.; Lasic  D. D.; and Israelachvili, J.
N., Biophys. J. {\bf 1994}, 66, 1479.


\bibitem{nano} Mangeney, C. ; Ferrage, F.; Aujard I. et al, J. Am. Chem. Soc.,
{\bf 2002}, 124, 5811.

\bibitem{ref19} Phan, S. ; Russel,  W. B.; Cheng, Z.; Zhu, J.; Chaikin, P. M.;
Dunsmuir J. H.; and Ottewill, R.H., Phys. Rev. E {\bf 1996}, 54, 6633.

\bibitem{ref18} Quirantes, A.; and  Delgado, A. V., J.Phys.D: Appl. Phys. 
{\bf 1997}, 30, 2123.

\bibitem{ref7} de Gennes, P. G., J. Phys.(Paris), {\bf 1976}, 37, 1443; 
Macromolecules, {\bf 1980}, 13, 1069; C. R. Acad. Sci. (Paris), {\bf 1985}, 300,
839.

\bibitem{ref8} Alexander, S., J. Phys.(Paris), {\bf 1977}, 38, 983.

\bibitem{ref9} Semenov, A. N., Sov. Phys. JETP, {\bf 1985}, 61, 733.

\bibitem{ref10}Milner,  S. T.; Witten T. A.; and Cates M. E.,  Macromolecules
{\bf 1998}, 21, 2610; Europhys. Lett., {\bf 1988}, 5, 413; Macromolecules
{\bf 1989}, 22, 853.

\bibitem{ref11}  Milner, S. T.; and Witten, T. A., J. Phys.(Paris), {\bf 1998}, 49,
1951.

\bibitem {zhu}Zhulina, E. B.; Borisov  O. V.; and Priamitsyn, V. A., J. Colloid
Interface Sci., {\bf 1990}, 137, 495.

\bibitem{ref12} Chakrabarti, A.; and Toral R., Macromolecules, {\bf 1990}, 23,
2016.

\bibitem{ref13} Lai, P.Y.; and Binder, K., J. Chem. Phys., {\bf 1991}, 95, 9288.
 
\bibitem{ref14} Toral,  R.; Chakrabarti, A.; and Dickman,  R.,
Phys. Rev. E, {\bf 1994}, 50, 343.  

\bibitem{ref15} Chakrabarti, A.; Nelson P.; and Toral, R., Phys. Rev. A,{\bf 1992},
46, 4930; J. Chem. Phys. {\bf 1994}, 100, 748.

\bibitem{ref16} Murat, M.; and Grest, G.S., Macromolecules, {\bf 1989},22, 4054;
Phys. Rev. Lett. {\bf 1989}, 63, 1074.

\bibitem{8}  Semenov, A. N.; Sov. Phys. JETP, {\bf 61}, 733, (1985) (Zh. Eksp.
Teor. Fiz., {\bf1985}, 88, 1242.

\bibitem{grant} Grant, M. C.; Russel, W. B., Phys. Rev. E, {\bf 1993}, 47, 2606.

\bibitem{ref25} Nommensen,  M. H.; Duits, G.; van den Ende, D.; and Mellema, J.,
Phys. Rev. E, {\bf 1999}, 59, 3147. 

\bibitem{6c} Cosgrove, T.; Crowley,  T.L; Vincent, B.; Barnett K.G.; and 
Tadros, Th. F., Faraday Discussions,{\bf 1982}, 16, 101.

\bibitem{6d}Beaufils, J.P.; Hennion, M.C.; and Rosset, R., J. de Physique,{\bf
1983}, 44, 497.

\bibitem{2k} Hommel, H.; Legrand, A.P; Balard,  H.; and Papirer, E., Polymer,{\bf 1983},
24, 959.

\bibitem{6f} Hommel, H.; Legrand, A.P.,  J. le Courtier and J. Desbarres, Eur.
Polym. J.,{\bf 1979}, 15, 993.

\bibitem{6g}Gilpin, D. K.; and Geindoga, E., J. Chromatogr. Sci.,{\bf 1983},
21, 352.

\bibitem{6h}Cosgrove, T.; Vincent, B.; Stuart, M. C.; Barnett, K.G.; and
Sissons,  D.S., Macromolecules, {\bf 1981}, 14, 1018.

\bibitem{ref20}  Daoud, M.; and Cotton, J. P., J. Physique, {\bf 1982}, 43, 531.

\bibitem{ref21} Witten, T. A.; and Pincus, P. A., Macromolecules, {\bf 1986}, 19,
2509.

\bibitem{ref22} Borukhov, I.; and Leibler, L., Phys. Rev. E, {\bf 2000}, 62, R41.

\bibitem{25}  Dan, N.; and Tirell, M., Macromolecules,{\bf 1992}, 25,2890.

\bibitem{30}  Toral, R.; and Chakrabarti, A., Phys. Rev. E, {\bf 1993},
47 ,4240.

\bibitem{30a} Grest,G.S.; and Murat, M., Macromolecules {\bf 1993}, 26, 3108. 


\bibitem{Wijmans} Wijmans, C.M.; Leermakers, F.A.M.; and Fleer, G.J., Langmuir
{\bf 1994}, 10, 4514. 

\bibitem{CST} Cerd\`{a}, J. J.; Sintes, T.; and Toral, R., Macromolecules {\bf 2003},
36, 1407.

\bibitem{29} Baung\"artner, A., in {\em Applications of the Monte Carlo Method
in Statistical Physics}, 2$^{nd}$ ed. Edited by K. Binder. Topics in Current
Physics 36. Springer-Verlag, Berlin, 1987.

\bibitem{link}Allen,  M.; and Tildesley, D., {\em Computer Simulation of Liquids};
Clarendon, Oxford, 1987.

\bibitem{7e} de Gennes, P. G.,{\em Scaling concepts in polymer physics};
Cornell University,Ithaca NY, 1979.

\bibitem{4a} Edwards, S.F., Proc. Phys. Soc.,{\bf 1965},85, 613.

\bibitem{Carig} Carignano, M.A.; and Szleifer, I., J. Chem. Phys, {\bf 1995}, 102,
8662.

\bibitem{4h} Flory, P. J. {\em Principles of Polymer Chemistry}; Cornell
University Press, 1953.

\bibitem{3e}Cosgrove, T.; Heath, T.; van Lent, B.; Leermakers, F.A.M.; and 
Scheutjens, J. M. H. M., Macromolecules,{\bf 1987}, 20, 1692. 


\bibitem{7c}Ruckenstein,  E.; and Li, B., J.Chem.Phys {\bf 1997}, 107, 3.

\bibitem{4f} Lin, E.K.; and Gast, A.P., Macromolecules,{\bf 1996},29, 390.

\bibitem{EvNa} Evans, R.; and Napper,  D. H., J.Colloid Interface Sci, {\bf 1978}, 63,
43.


\end{thebibliography}
\end{document}